\documentclass[aps,rsi,twocolumn,pre,nofootinbib]{revtex4-2}
\usepackage{graphicx}
\usepackage{epstopdf, epsfig}
\usepackage{newtxtext}
\usepackage{newtxmath}
\usepackage{natbib}
\usepackage{hyperref}
\usepackage[T1]{fontenc}
\usepackage{bm}
\usepackage{amsmath}
\usepackage{color}
\usepackage{xcolor, colortbl}

\newcommand{\beq}{\begin{equation}}
\newcommand{\eeq}{\end{equation}}
\newcommand{\ba}{\begin{array}}
\newcommand{\ea}{\end{array}}
\newcommand{\bee}{\begin{eqnarray}}
\newcommand{\eee}{\end{eqnarray}}

\makeatletter
\begin{document}

\title{Scaling law for a buckled elastic filament in a~shear flow}
\author{Pawe\l\ Sznajder}\thanks{Pawe{\l} Sznajder and Piotr Zdybel contributed equally to this work.}
\author{Piotr Zdybel$^*$}
\author{Lujia Liu}\thanks{Present address:  Research Institute of Aero-Engine, Beihang University, Beijing 100191, PR China.}
\author{Maria~L.~Ekiel-Je\.zewska}\email{pzdybel@ippt.pan.pl, mekiel@ippt.pan.pl}

\affiliation{Institute of Fundamental Technological Research, Polish Academy of Sciences, Pawi\'nskiego 5B, 02-106 Warsaw, Poland}

\date{\today}
\begin{abstract}
We analyze the three-dimensional buckling of an elastic filament in a shear flow of a viscous fluid at low Reynolds number and high Peclet number. We apply the Euler-Bernoulli beam (elastica) theoretical model. We show the universal character of the full 3D spectral problem for the small perturbation of the thin filament from a straight position of arbitrary orientation. We use the eigenvalues and eigenfunctions for the linearized elastica equation in the shear plane, found earlier by [Liu et al., 2024] with the Chebyshev spectral collocation method, to solve the full 3D eigenproblem.  We provide a simple analytic approximation to the eigenfunctions,  represented as Gaussian wavepackets. As the main result of the paper, we derive square-root dependence of the eigenfunction wavenumber on the parameter 
$\tilde{\chi}=-\eta \sin 2\phi \sin^2\theta$, where $\eta$ is the elastoviscous number, and the filament orientation is determined by the zenith angle $\theta$ with respect to the vorticity direction and the azimuthal angle $\phi$ relative to the flow direction.
We also compare the eigenfunctions with shapes of slightly buckled elastic filaments with a non-negligible thickness with the same Young's modulus, using the bead model and 
performing numerical simulations with the precise {\sc Hydromultipole} numerical codes.
\end{abstract}
\maketitle


\section{Introduction}
Dynamics of elastic filaments in low-Reynolds-number fluid flows have been recently intensively investigated \cite{shelley2011,Linder2015,du2019dynamics,diamant2020}. The growing interest is related to the progress of the experimental techniques to trace the behavior of biological micro-objects such as flagella of bacteria or algae, chains of diatoms, actin, or cilia \cite{berg1973,review2006flagellar,Goldstein2009,musielak,rafai,kantsler2012,harasim2013,liu2018,diatoms2020}. Moreover, the development of modern technology has led to the production of elastic nano- and microfibers with controlled length, width, and Young's modulus \cite{reneker2008electrospinning,nunes2013,nakielski2020} with wide perspectives for their applications. 

Therefore, the motion and typical shape deformations of elastic fibers have been studied in various fluid flows \cite{yamamoto1993,Skjetne1997,Tornberg2004,autrusson2011,farutin,liu2018,Nguyen2014,lagrone2019complex,zuk2021universal,Kurzthaler2023}, with a special interest in the 
buckling instability~\cite{young2007stretch,wandersman2010,guglielmini2012,kantsler2012,manikantan2015,quennouz2015,hall,kanchan2019,marchioli2023slender}. In particular, the buckling of elastic filaments in a shear flow at low-Reynolds-number has been extensively investigated in the literature. In Ref.~\cite{Forgacs1959a,forgacs1959particle}, experiments were performed and the minimum value of the shear rate at which thin filaments can buckle was evaluated from the Euler equation. 
In ref.~\cite{Becker2001}, the buckling instability of thin elastic filaments was demonstrated by solving the spectral problem for the Euler-Bernoulli beam confined to the shear plane (i.e., the plane spanned by the flow and the flow gradient directions). In Ref.~\cite{liu2022}, 
the spectral problem 
with perturbations also out of the shear plane was analyzed and it was shown that it is the same as in the compressional flow, studied earlier in Ref.~\cite{chakrabarti2020}. 
In Ref.~\cite{slowicka2022},
the 3-dimensional buckling of thick fibers in a shear flow was studied experimentally and numerically, and it was shown that the buckling is limited to short times only.

In this work, we investigate the buckling instability of an elastic slender filament in the shear flow. We linearize the elastica model~\cite{Audoly2000,Becker2001} around an arbitrary orientation, 
solve the full 3-dimensional spectral problem, and derive a scaling law for the eigenvalues and eigenfunctions. We also compare the results with our numerical simulations of flexible fibers of a non-zero thickness.  

The plan of the paper is the following. In Sec.~\ref{sec:elastica}, the system is introduced. The elastica equation for a slender elastic filament in the shear flow is linearized. The fully three-dimensional spectral problem for the perturbation growth is reduced to solving a single equation with just one parameter $\tilde{\chi}$. The eigenvalues and eigenfunctions are presented. Sec.~\ref{sec:approximation} contains a simple analytical approximation to the eigenfunctions as the Gaussian wavepackets, valid for large values of $\tilde{\chi}$ (i.e., for highly flexible filaments). Within this approximation, it is shown that the characteristic wavenumber of the most unstable eigenfunctions scales as $\sqrt{\tilde{\chi}}$.  
  In Sec.~\ref{sec:comparing}, the dependence of the eigenfunctions on $\tilde{\chi}$ is analyzed. 
The fast Fourier transform of the filament local curvature is evaluated for the elastica most unstable eigenfunctions. 
Conclusions are presented in Sec.~\ref{sec:conc}. In Appendix~\ref{A1},  the asymptotic expansion is applied to provide the scaling of the eigenvalues and eigenfunctions in a range of very small values of the arclength $s$. In Appendix~\ref{A2}, the difference between the wavenumber based on shape and its curvature is estimated. 
In Appendix~\ref{sec:bead_model}, the thickness of an elastic filament is taken into account within the bead model and the multipole method, implemented in the {\sc Hydromultipole} numerical codes. The numerical simulations are performed and the shapes of the slightly buckled filament are shown and compared with the elastica eigenfunctions.
 
\section{Euler-Bernoulli beam theory for an elastic filament in a shear flow}\label{sec:elastica}

\subsection{Stability of a small 3D perturbation from a straight configuration}

We consider an elastic filament of length $L$ and circular cross-section of diameter $d$ under compression of a shear flow 
\beq\bm{v}_0=(\dot{\gamma} y, 0, 0)\label{shear}\eeq
of a viscous fluid with the dynamic viscosity $\mu$. The Reynolds number  Re $\ll 1$, and the Peclet number Pe $\gg 1$. We assume that the filament is infinitely thin, which allows describing its evolution with the use of the Euler-Bernoulli beam theory, as in \cite{Audoly2000,Audoly2015,Becker2001,Linder2015,liu2018,chakrabarti2020,liu2022}. The dimensionless evolution equation, based on the slender body hydrodynamic interaction proposed by \cite{batchelor1970slender}, reads 
\begin{equation}
    \eta (2 \bm{I}-\bm{x}_s\bm{x}_s)\cdot (\bm{x}_t-\bm{U}(\bm{x}))=(T(s,t)\bm{x}_s)_s-\bm{x}_{ssss},
    \label{main}
\end{equation}
where $s\in [-1/2,1/2]$ is the arclength along the filament, $\bm{x}(s,t)$ is the position of a filament segment (both normalized by the filament length $L$), $t$ is time normalized by $1/\dot{\gamma}$, $\bm{U}=\bm{v}_0/(\dot{\gamma}L)$,
$T(s,t)$ is the tension, the filament is inextensible (i.e., $\bm{x}_s \cdot \bm{x}_s=1)$ and 
\bee
\eta=\frac{2\pi \mu \Dot{\gamma}L^4}{EI \ln(2L/d)}\label{def eta}
\eee
is the elastoviscous number, where $E$ is Young's modulus and $I=\pi d^4/64$ is the area moment of inertia, see  \cite{Becker2001}.

\begin{figure}[h!tbp] \vspace{-0.8cm}
\centerline{\includegraphics[width=0.7\textwidth]{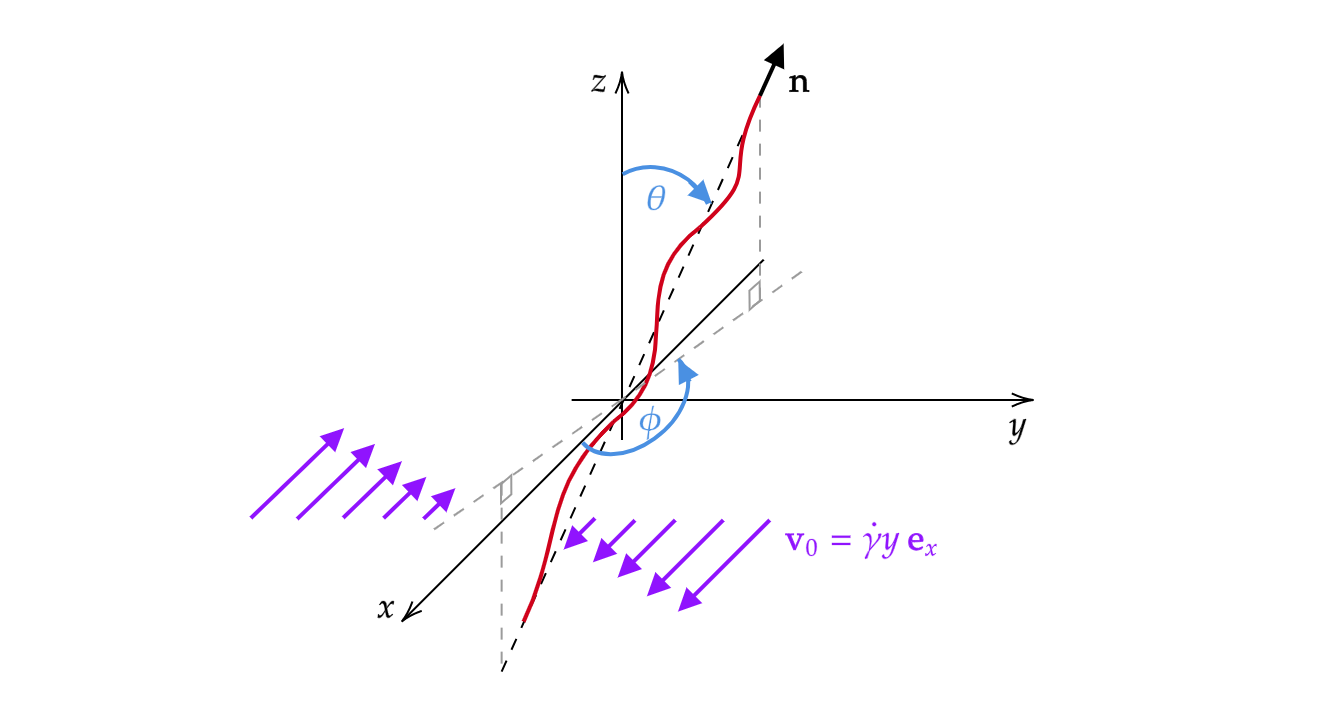}}\vspace{-0.5cm}
	\caption{A schematic of a slender fiber in a shear flow.}\label{sch}
\end{figure}

As illustrated in Fig.~\ref{sch}, the fiber is almost straight, at an arbitrary orientation $\bm{n}$, determined by the spherical angles $0<\theta<\pi$ and $\pi/2<\phi<\pi$ (with the zenith axis along $z$). There are small perturbations $u$ and $v$ along two unit vectors perpendicular to $\bm{n}$ and to each other: $\bm{\lambda}_i$ in the shear plane $xy$ and $\bm{\lambda}_o$ out of the shear plane, respectively. Therefore, $\bm{x}=s\bm{n}+u\bm{\lambda}_i+v\bm{\lambda}_o$. We assume the free boundary conditions at the filament ends, 
\beq u_{ss}=u_{sss}=v_{ss}=v_{sss}=0 \mbox{ at } s=\pm \frac{1}{2}.
\eeq

To study the stability of the perturbations, we linearize the elastica equation \eqref{main} around the straight shape at the orientation $\bm{n}$, arriving at a system of coupled differential equations for $u$ and $v$, as in Ref.~\cite{liu2022}. Then we use the spectral method for an arbitrary orientation $\bm{n}$. We assume that \beq
[u, v]=[\Phi_u(s), \Phi_v(s)]\exp(\sigma t),\eeq with the real in-plane and out-of-plane shapes,
$\Phi_u(s)$ and $\Phi_v(s)$, and the complex number $\sigma$.
From the linearized equation \eqref{main},  we derive the following 3D set of coupled equations for the eigenfunctions $\Phi_u(s)$ and $ \Phi_v(s)$, 
\bee
2\eta\sigma \begin{bmatrix} \Phi_u \\ \Phi_v \end{bmatrix}\! &=& \!\begin{bmatrix} \alpha\hat{\bm{I}}+\hat{\mathcal{L}} & 0 \\ \beta & \alpha'\hat{\bm{I}}+\hat{\mathcal{L}} \end{bmatrix}  \begin{bmatrix} \Phi_u \\ \Phi_v \end{bmatrix}\nonumber \\
&=&\begin{bmatrix} \alpha\Phi_u+\hat{\mathcal{L}}\Phi_u \\ \beta\Phi_u +\alpha'\Phi_v+\hat{\mathcal{L}}\Phi_v \end{bmatrix},\label{full3D}\\
    \hat{\mathcal{L}}(\tilde\chi)&=&-\frac{\partial^4}{\partial s^4}+\tilde\chi\left[\frac{1}{4}(s^2-\frac{1}{4}) \frac{\partial^2}{\partial s^2}+s \frac{\partial}{\partial s}\right],\hspace{0.7cm}\label{L}
\eee
with the identity operator $\hat{\bm{I}}$, and 
$\tilde\chi\!=\!-\eta\sin(2\phi)\sin^2 \theta$,
$\alpha\!=\!-\eta\sin(2\phi)$, $\alpha'\!=\!\eta\sin(2\phi)\cos^2\theta$, 
$\beta\!=\!-2\eta \cos(2\phi)\cos\theta$.

\subsection{Universal character of the spectrum and the eigenfunctions}

Before giving a general solution to Eqs~\eqref{full3D}-\eqref{L}, we first consider the special case with  $\beta=0$. It corresponds to $\theta=\pi/2$ (i.e., the filament in the shear plane $xy$, also discussed in Ref.~\cite{liu2022}), or $\phi=\pi/4$ or $3\pi/4$. 
For $\beta=0$,  the in-plane, and out-of-plane perturbations are not coupled, and the set of Eqs. \eqref{full3D} has the form 
\begin{equation}
 2\eta\begin{bmatrix} \sigma_{u1} \Phi_{u1} \\ \sigma_{v1} \Phi_{v1} \end{bmatrix} 
=\begin{bmatrix} \alpha\Phi_{u1} +\hat{\mathcal{L}}\Phi_{u1} \\ \alpha'\Phi_{v1} +\hat{\mathcal{L}}\Phi_{v1} \end{bmatrix},\label{coupled}
\end{equation}
with  separate eigenvalues $\sigma_{u1}$ and $\sigma_{v1}$, and separate eigenfunctions $\Phi_{u1}$ and $\Phi_{v1}$.

We recall that after adding to an operator $\hat{\mathcal{L}}$ the identity operator multiplied by any 
number $\gamma$, the eigenvectors $\Phi_1$ remain the same, and  $\gamma$ is added to the eigenvalues $\lambda$\footnote{By definition, $\Phi_1$ is an eigenvector of an operator $\mathcal{L}$ for an eigenvalue $\lambda$ iff $\mathcal{L}\Phi_1=\lambda \Phi_1$.}, 
\begin{equation}
    (\hat{\mathcal{L}}+\gamma \hat{\bm{I}} )\Phi_1=
    (\lambda 
    +\gamma)\Phi_1.
\end{equation}
Therefore, we can set $\Phi_{v1}=\Phi_{u1}$ and then Eq.~\eqref{coupled} can be rewritten as
\beq (2\eta\sigma_{u1}-\alpha)\Phi_{u1}=\hat{\mathcal{L}}\Phi_{u1}= (2\eta \sigma_{v1} -\alpha')\Phi_{u1}.\label{eqegv}\eeq

Summarizing: for $\beta=0$, the sets of the in-plane and out-of-plane eigenfunctions are the same, and the in-plane and out-of-plane eigenvalues are shifted concerning each other by $\alpha'-\alpha$,
\bee \sigma_{v1}=\left\{ \ba{l}
\sigma_{u1}+\frac{\sin(2\phi)}{2}\mbox{ for }\theta=\pi/2,
\\
\sigma_{u1}\pm \frac{(\cos^2\!\theta +1)}{2}\mbox{ for }\phi=\pi/4, \, 3\pi/4.
\ea\right. 
\eee

For the unperturbed filament in the shear plane $xy$, i.e., for $\theta=\pi/2$, the eigenproblem has been discussed in Ref.~\cite{liu2022}. It has been also pointed out there that for $\theta=\pi/2$, the eigenproblem in the shear flow, given by Eqs \eqref{coupled}, is the same as in Ref.~\cite{chakrabarti2020} for the compressional flow, is the parameters are matched to each other accordingly. 

However, the full 3D eigenproblem \eqref{full3D}-\eqref{L}, with the coupling between the in-plane and out-of-plane perturbations (i.e., for $\beta \ne 0$), has not been studied so far.

Therefore, now we move on to the 3D system of Eqs \eqref{full3D} for $\beta \ne 0$. In this case, Eqs \eqref{full3D}  are coupled with each other.  Moreover, the equations for the in-plane perturbation in the systems of Eqs \eqref{coupled} and \eqref{full3D} are identical, and therefore $\sigma \equiv \sigma_{u1}$ and $\Phi_{u}=\Phi_{u1}$. Then we  check if there exists such  $\delta$ that $\Phi_{u}$ and  $\Phi_{v}=\delta \Phi_{u}$ are the solution of 
Eqs \eqref{full3D}, which, after applying Eq. \eqref{eqegv}, take the form
\begin{equation}
2\eta\sigma_{u1} \begin{bmatrix} \Phi_{u1} \\ \delta \Phi_{u1} \end{bmatrix} 
=\begin{bmatrix} 
(\alpha+\hat{\mathcal{L}})\Phi_{u1} \\ \beta\Phi_{u1} +2\eta\sigma_{v1}\delta\Phi_{u1} \end{bmatrix}.\label{bn0}
\end{equation} 
The second equation gives the following result for $\delta$, 
\begin{equation}
    \delta=\frac{\beta}{2\eta\sigma_{u1}-2\eta\sigma_{v1}}=\frac{\beta}{\alpha-\alpha'}
 =\frac{2\cos\theta}{\tan(2\phi)(1+\cos^2\theta)}.\label{delta_of_theta}
\end{equation}
Therefore, for a given eigenvalue $2 \eta \sigma$ of the operator $\alpha \hat{\bm{I}} + \hat{\mathcal{L}}$, the shape of the out-of-plane eigenfunction $\Phi_{v}$ is the same as the shape of the in-plane eigenfunction $\Phi_{u}$ with the rescaled amplitude.

\subsection{The basic spectral problem}\label{spectralproblem}
We now point out that with the explicit form \eqref{L} of the operator $\hat{\mathcal{L}}$, the spectral  equation \eqref{bn0} depends only on a single parameter $\tilde{\chi}$,
\begin{equation}
    \left[-\frac{\partial^4}{\partial s^4}+\tilde{\chi}\left[\frac{1}{4}(s^2-\frac{1}{4}) \frac{\partial^2}{\partial s^2}+s \frac{\partial}{\partial s}\right]-\tilde{\chi}\tilde{\sigma}\right]\Phi_{u1}=0,\label{unieq}
\end{equation}
where
\beq
\tilde{\sigma}=\frac{-2(\sigma +\sin(2\phi)/2)}{\sin(2\phi)\sin^2\theta}.\label{si}
\eeq

This equation is identical to the spectral equation for the elastic fiber in a shear flow with the in-plane perturbations only, solved in Ref.~\cite{Becker2001}, and the spectral equation for the elastic fiber in the compressional flow, solved in Ref.~\cite{chakrabarti2020}, if the parameters are matched accordingly.

The eigenproblem \eqref{unieq}-\eqref{si} has been solved by the Chebyshev collocation method, 
as described in Ref.~\cite{liu2022,repository_2023}.
\begin{figure}[h!tbp]
\centerline{\includegraphics[width=0.45\textwidth]{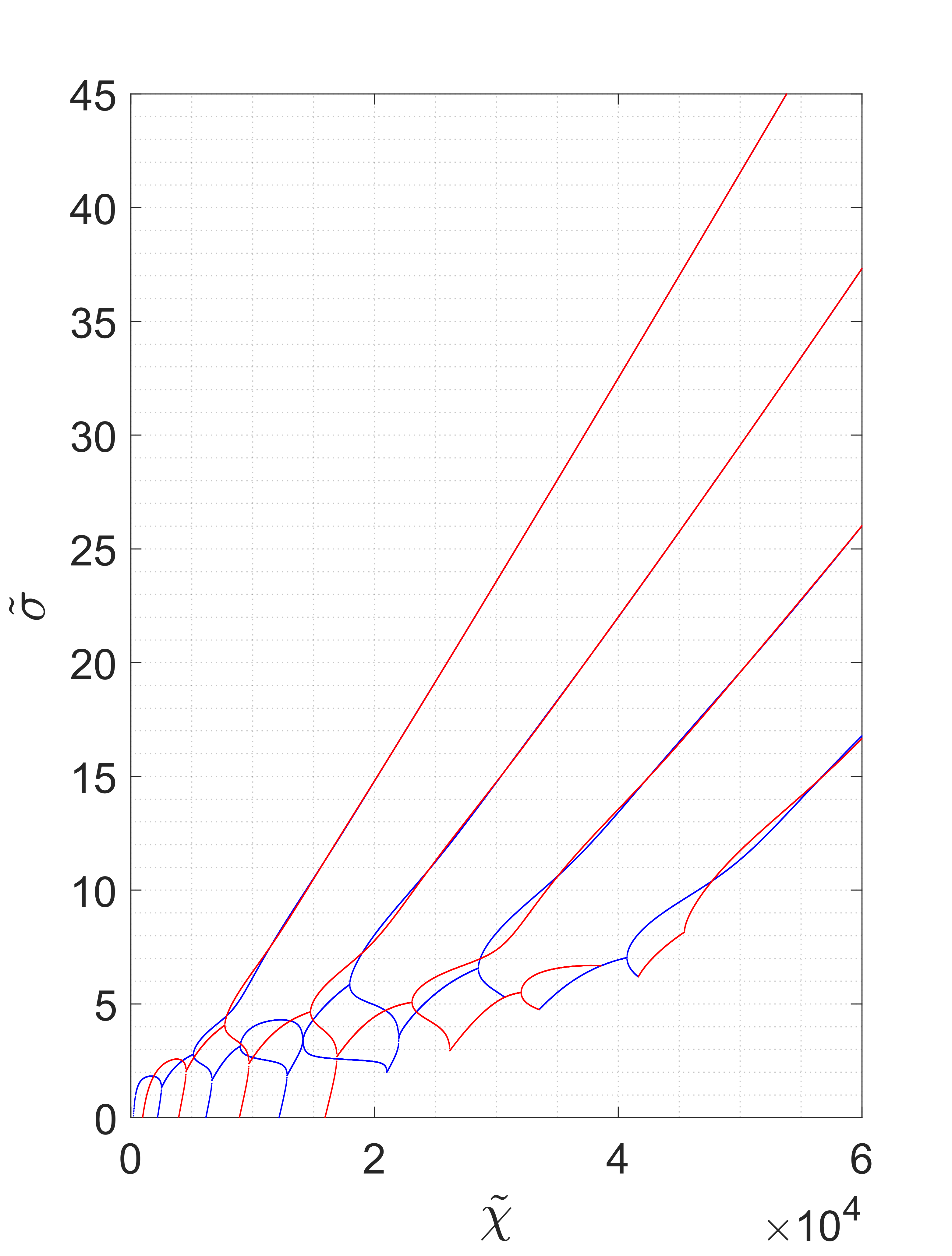}}\vspace{-0.4cm}
	\caption{Solution of the spectral problem \eqref{unieq}, taking into account no more than 8 most unstable eigenvalues. Eigenvalues for odd (red) and even (blue) eigenfunctions.}
 \label{spectfigI}
\end{figure}
\begin{figure}[h!tbp]
\centerline{\includegraphics[width=0.45\textwidth]{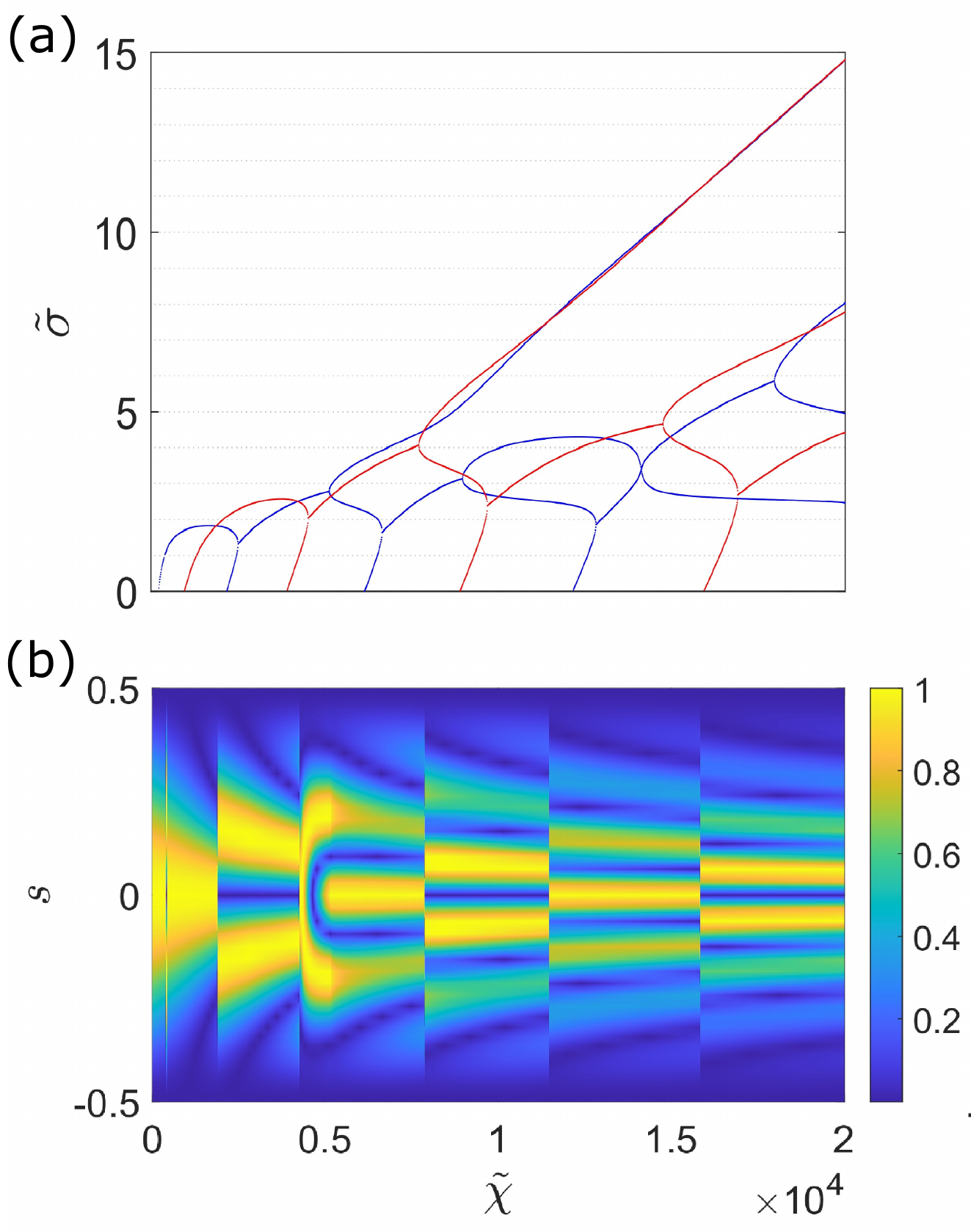}}\vspace{-0.4cm}
	\caption{(a). An enlargement of Fig.~\ref{spectfigI}. (b). The most unstable eigenfunctions. The colors indicate $\kappa(s)/\max_s\kappa(s)$, the local curvature of $\Phi_{u1}(s)$. The range of $\tilde{\chi}$ in both panels is the same.}
 \label{spectfigII}
\end{figure}
The dependence of a few consecutive larger (including the largest) values of $\tilde{\sigma}$ on $\tilde{\chi}$ is shown in Fig.~\ref{spectfigI}, with an enlargement at the  Fig.~\ref{spectfigII}(a). Red and blue colors denote odd and even eigenfunctions, respectively.  (Compare with the eigenvalues presented in Refs.~\cite{Becker2001,chakrabarti2020,liu2022}.) 
\begin{figure*}[t!]
	\centerline{\includegraphics[width=0.9\textwidth]{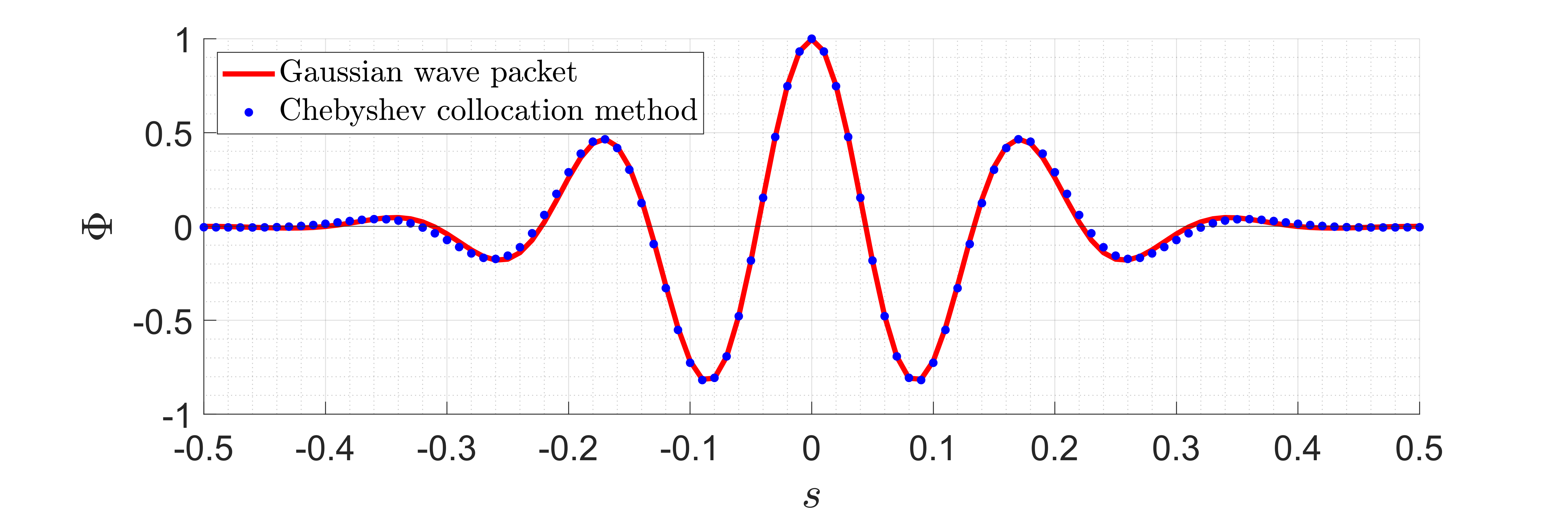}}\vspace{-.4cm}
	\caption{ The eigenfunction $\Phi$ as a function of $s$. Here $\Phi$  corresponds to the most unstable eigenvalue for $\tilde{\chi}=5.4\times 10^4$. Data depicted by blue dots are computed using the Chebyshev collocation method, and the red line displays the fitted shape of the Gaussian wave packet. }
	\label{ksztalt_modu_fit}
\end{figure*}

In Fig.~\ref{spectfigII}(b), the local curvature of the eigenfunction corresponding to the largest $\tilde{\sigma}$ is also presented with the use of colors. The local curvature is defined in the standard way, approximated for small deformations as $\kappa(s)\!=\!|\tilde{\Phi}_{u1}''(s)|/\left(1+\left(\tilde{\Phi}_{u1}'(s)\right)^2\right)^{3/2}\!\!\!\approx\! |\tilde{\Phi}_{u1}''(s)|$, and normalized by $\max_s \kappa(s)$, where $\tilde{\Phi}(s)=\Phi (s)\times 10^{-2}$. Discontinuities, visible in Fig.~\ref{spectfigII}(b), appear at the same values of $\tilde{\chi}$ for which blue and red lines cross each other in the Fig.~\ref{spectfigII}(a), which means that parity of the most unstable eigenfunction is changed.  Such a coupling of the largest odd and the largest even eigenfunctions motivates us to investigate in Sec.~\ref{scalinghere} both of them as continuous functions of $\tilde{\chi}$, similarly as done in Ref.~\cite{chakrabarti2020} for the compressional flow.  

Fig. \ref{spectfigI} illustrates that in most cases, the last and the last but one eigenvalue correspond to eigenfunctions of a different parity, and therefore the spectrum degenerates at their crossing point. However, 
sometimes it happens that the last and the last but one eigenvalue correspond to eigenfunctions of the same parity, and they tend to coincide when $\tilde{\chi}$ goes to a certain limiting value  (e.g., such a branching point is visible in Fig.~\ref{spectfigI} at $\tilde{\chi}\approx 5130$).
Those values of $0\le\tilde{\chi}\le 10^4$ which are the crossing or branching points described above,  
correspond to the borders between consecutive "modes" shown in  Fig. 2 of Ref.~\cite{Becker2001} for elastica in the shear flow. 
\begin{figure*}
\centerline{\includegraphics[width=0.95\textwidth]{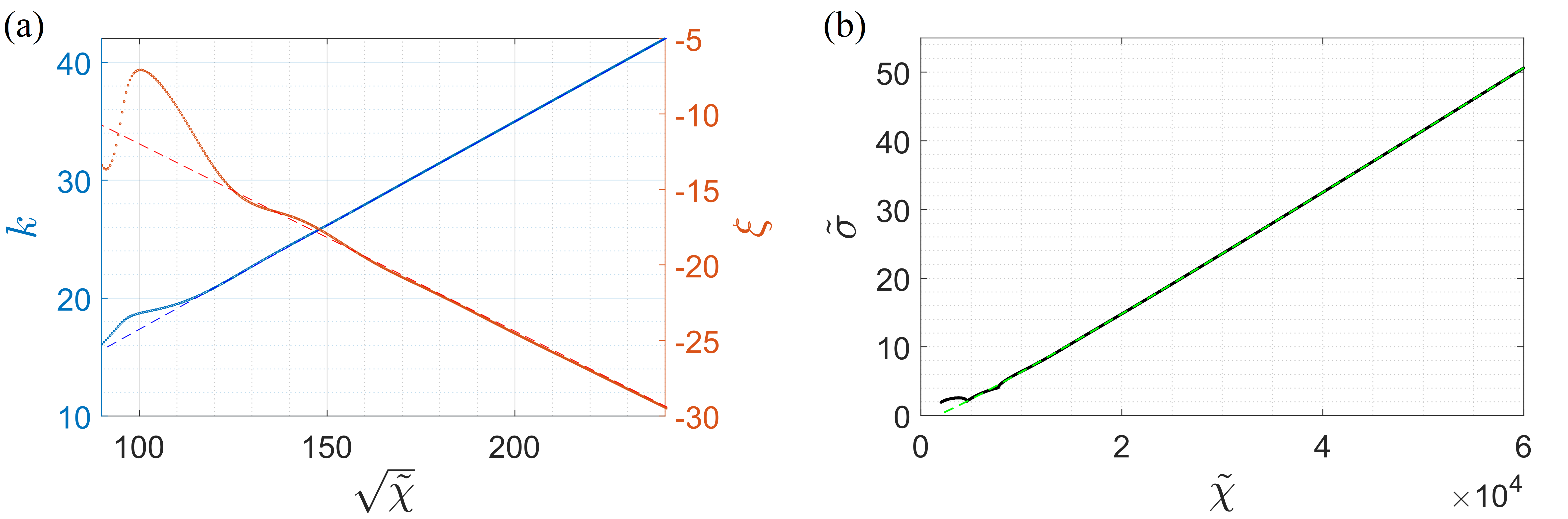}}\vspace{-0.3cm}
	\caption{Scaling of $k$, $\xi$ and $\tilde{\sigma}$
 with $\tilde{\chi}$. Dots: parameters of the Gaussian wave packet approximation to the eigenfunctions with the most unstable eigenvalue $\tilde{\sigma}$, determined numerically from the Chebyshev collocation method. Dashed lines: fits from Eqs. \eqref{fitk}-\eqref{fitsigma}, respectively.
 (a). The wavenumber $k$ (blue) and the measure of dispersion $\xi$ (red)   vs.  $\tilde{\chi}^{1/2}$. 
(b).~Dependence of the most unstable eigenvalue $\tilde{\sigma}$  on  $\tilde{\chi}$ (black dots and green dashed line).}
	\label{k_mu_spect}
\end{figure*}

\section{Approximate analytical solution at large values of \texorpdfstring{$\tilde{\chi}$}{Lg}}\label{sec:approximation}
In this section, we propose a simple analytical approximation of the most unstable eigenvalues and eigenfunctions in Eq.~\eqref{unieq} for elastica in a shear flow. We assume that 
a~hydrodynamic forcing dominates an internal elasticity of a~filament, and the elastoviscous number $\eta$ is large. We restrict to such angles $\theta$ and $\phi$  that 
 $\tilde\chi\!=\!-\eta\sin(2\phi)\sin^2 \theta$ is also large. We focus on $\tilde{\chi} \ge 1.5\cdot 10^4$.
 
\subsection{Approximation of shape}
An example  of the 
most unstable eigenfunction acquired numerically by the Chebyshev spectral collocation method, as described in Ref.\cite{liu2022,repository_2023}, is presented in Fig.  \ref{ksztalt_modu_fit} (blue dots). The numerically received profile of the elastica
resembles the Gaussian wave packet. In general, eigenmodes $\Phi_{u1}$ are either even or odd functions of the argument $s$. Thus we propose  an approximate solution $\Phi_{u1} \approx \Phi$ in the form
\begin{equation}
    \Phi(s)=\sin(ks)\,\mathrm{e}^{\xi s^2} \mathrm{~~or~~} \Phi(s)=\cos(ks)\,\mathrm{e}^{\xi s^2},
    \label{post}
\end{equation}
with $\xi <0 $. This choice is convenient thanks to the evident physical interpretation of parameters $k$ and $\xi$ as a wavenumber and a measure of dispersion, respectively.  Above $k$ is normalized by $1/L$, and in turn, $\xi$ is measured in units of $1/L^2$, where $L$ is the length of the fiber. We fit the expressions \eqref{post} to the numerical data obtained via the spectral approach for a wide range of values of $\tilde{\chi}\le 6\cdot 10^4$. An example of the fit for $\tilde{\chi}=5.4\cdot 10^4$
is shown in Fig.~\ref{ksztalt_modu_fit} as the red line. Fig.~\ref{ksztalt_modu_fit} illustrates that discrepancies between the exact shape and its analytical approximation are minuscule, though perceptible close to the ends of the filament.  A similar agreement is observed for the whole range of values $1.5\cdot 10^4 \le \tilde{\chi}\le 6 \cdot 10^4$.

\subsection{Scaling of the most unstable eigenfunctions and eigenvalues 
with  \texorpdfstring{$\tilde{\chi}$}{Lg}}

Taking advantage of the aforementioned fitting procedure for the most unstable eigenfunctions,  we get $k$ and $\xi$ as a~function of $\tilde{\chi}$ using data obtained numerically from the Chebyshev collocation method (see \cite{liu2022,repository_2023}) for $\tilde{\chi}$ up to $6.0\times10^4$. We present the results in Fig. \ref{k_mu_spect}(a). We notice that for large values of $\tilde{\chi}$, both $k$ and $\xi$ show a linear dependence on $\tilde{\chi}^{1/2}$. The resulting fits for the range $\tilde{\chi}^{1/2}>160$ are given below:
\bee
&&k=K_1{\tilde{\chi}}^{1/2}+K_2,\label{fitk}\\
&&\xi=M_1{\tilde{\chi}}^{1/2}+M_2,
\label{fitmu}
\eee
with $K_1\!=\!0.176 \pm 0.001$, $K_2\!=\!-0.249 \pm 0.003$, 
$M_1\!=\!-0.124 \pm 0.001$ and $M_2\!=\!0.417 \pm 0.034$. 

In Fig. \ref{spectfigI}, the most unstable eigenvalue $\tilde{\sigma}$ seems to be linear in $\tilde{\chi}$ for large values of $\tilde{\chi}$. Further examination, illustrated in  Fig. \ref{k_mu_spect}(b), leads to the following fit for $\tilde{\chi}^{1/2}>200$:
\beq
\tilde{\sigma}=S_1\tilde{\chi}+S_2{\tilde{\chi}}^{1/2}+S_3,\label{fitsigma}
\eeq
where $S_1\!=\!(9.765 \pm 0.001)\times 10^{-4}$, $S_2\!=\!-(3.124\pm 0.004)\times 10^{-2}$ and $S_3\!=\!-0.314\pm 0.031$.

Motivation of the scaling \eqref{post}-\eqref{fitsigma} based on the asymptotic expansion is presented in Appendix~\ref{A1}. It is limited to values of the arclength $s$ close to the middle of the filament.

\section{The elastica eigenfunctions}
\label{sec:comparing}
\subsection{Local curvature}
To analyze the eigenfunctions for the linearized elastica equation, we use the local curvature $\kappa(s)/\max_s{\kappa(s)}$, 
\begin{figure}[h!tbp]\vspace{-0.2cm}
  \includegraphics[width=0.45\textwidth]{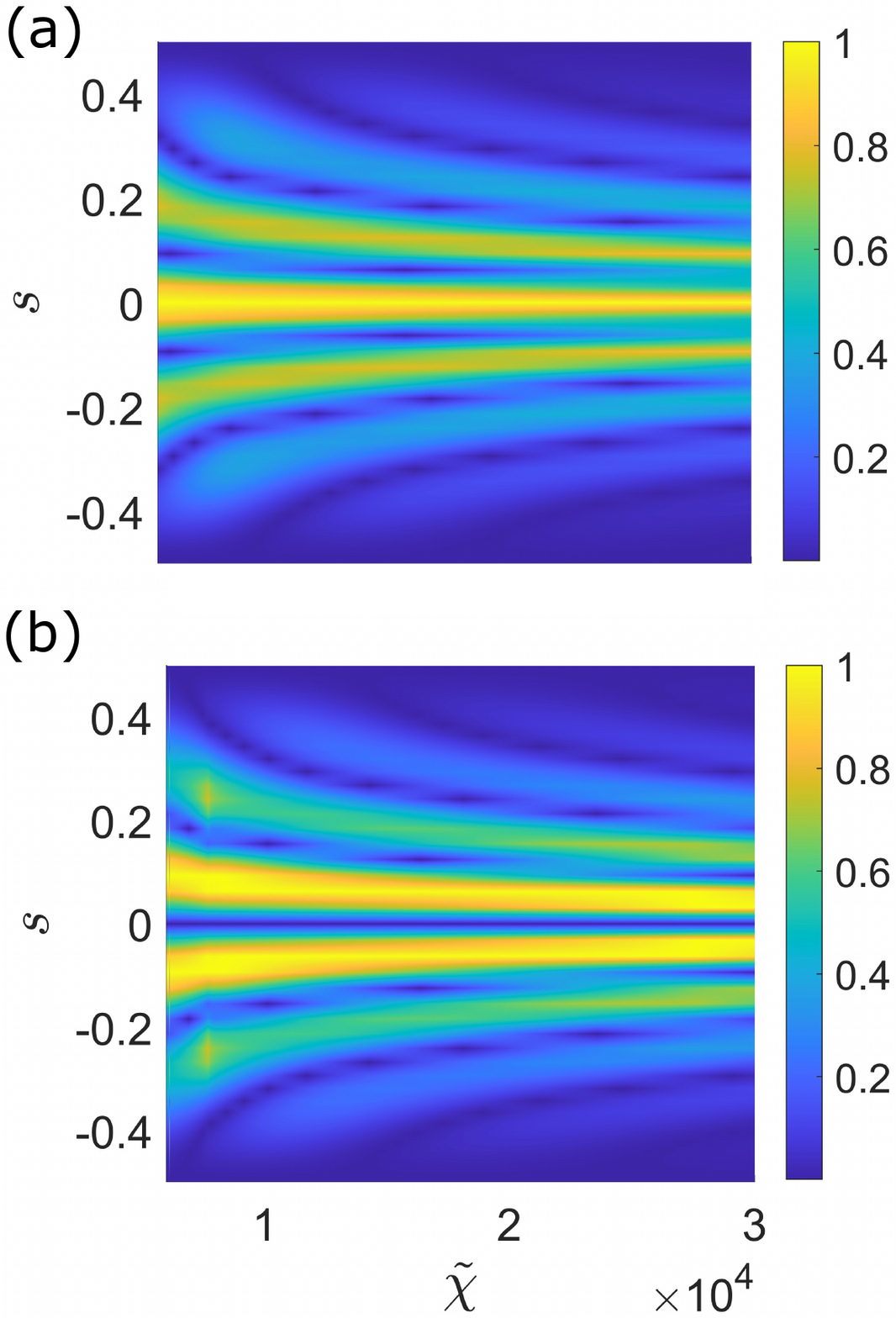}\vspace{-.6cm}
  \caption{Local curvature of an elastic filament, normalized by its maximum value along the filament and shown as colors. Theoretical results for the most unstable even (a), and odd (b) eigenfunctions for the linearized elastica equations.}\label{c4}
\end{figure}
 defined in Sec. \ref{spectralproblem}. The reason for this choice is that the local curvature is related to the fiber bending energy. 
 
 For a wide range of large values of $\tilde{\chi}$, the local curvature is shown in colors in Fig. \ref{c4}, separately for the even and odd most unstable eigenfunctions. 
We observe the increasing number of the local maxima for the increasing value of $\tilde{\chi}$, as in Fig.~12 from Ref.~\cite{slowicka2022}. This tendency will be analyzed more precisely in the next subsection.

\subsection{Scaling of the characteristic wavenumber}\label{scalinghere} 

\begin{figure}[h!tbp]
\includegraphics[width=0.45\textwidth]{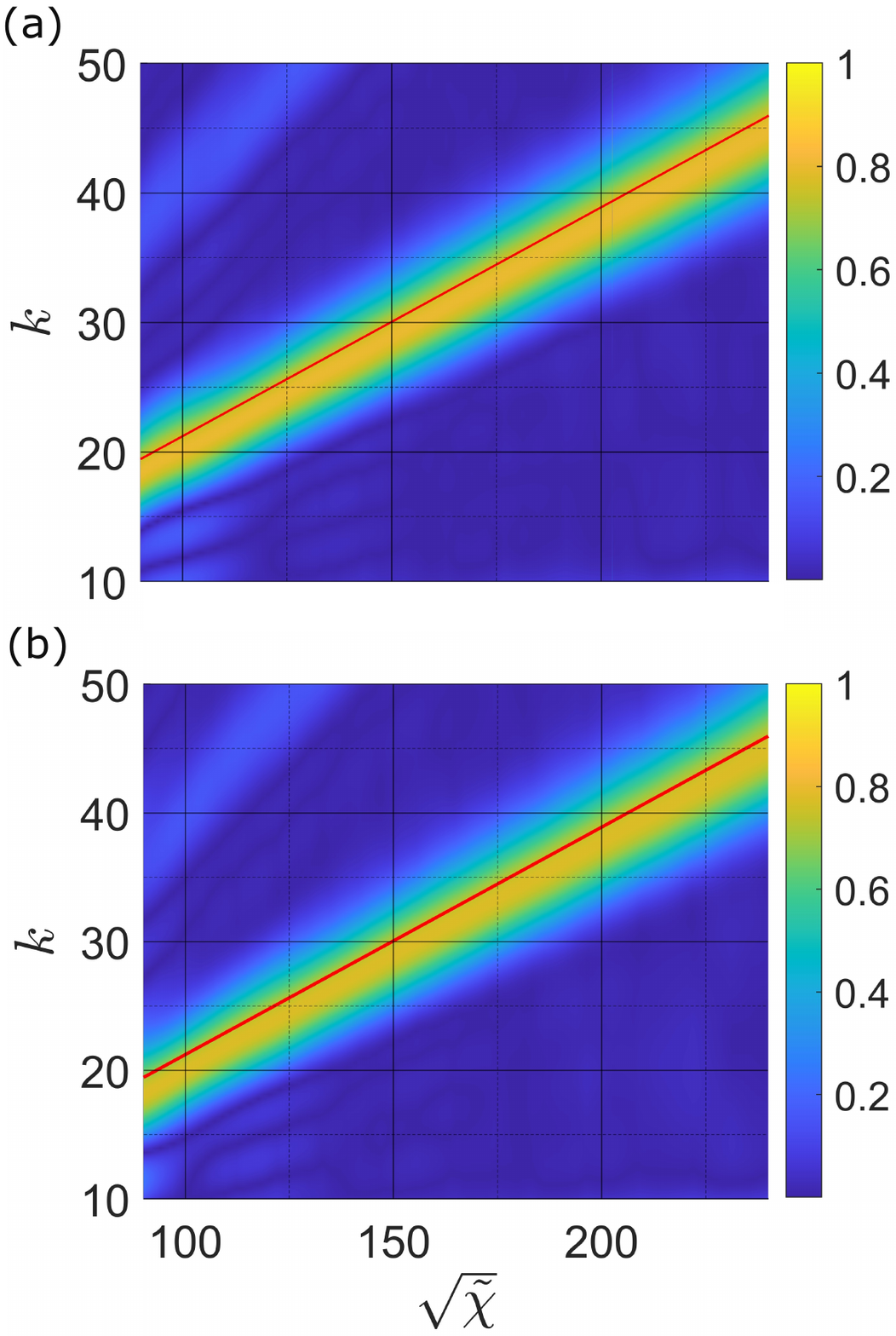}\vspace{-.4cm}
\caption{Wavenumber $k$ resulting from the fast Fourier transform (fft) of the local curvature of the elastica most unstable (a) even and (b) odd eigenfunctions vs. $\sqrt{\tilde{\chi}}$. The colors indicate the fft intensity, normalized by its maximum value along the filament. Red line: the scaling \eqref{alpha1} of the local curvature of the approximate Gaussian wavepacket, $\alpha_1\approx\frac{\sqrt{2}}{8}\sqrt{\tilde{\chi}}+\frac{5\sqrt{2}}{2}$ as a function of  $\sqrt{\tilde{\chi}}$.
}\label{k4}
\end{figure}

  We now perform the fast Fourier transform of the local curvature of the most unstable even and odd eigenfunctions, determined numerically by the Chebyshev spectral collocation method \cite{liu2022,repository_2023}. We plot the resulting wavenumber $k$ in Fig.~\ref{k4}, now as a function of $\sqrt{\tilde{\chi}}$, in a range of $\tilde{\chi} \gg 1$. The colors indicate the intensity of the fast Fourier transform (fft), divided by its maximum value.  
  For large values of $\tilde{\chi}$, we find that 
  $k$ is a linear function of $\sqrt{\tilde{\chi}}$. 

It seems straightforward to compare this linear relation with the dependence on $\sqrt{\tilde{\chi}}$ of the wavevector $k$ in the Gaussian wavepacket approximation to the most unstable even and odd eigenvalues, determined in Eqs.~\eqref{fitk} and \eqref{kms}.  
 However, we should keep in mind that there is a small difference between the dominant wavenumbers for $\kappa(s)$ and the corresponding eigenfunction $\Phi(s)$. 
This difference in Appendix~\ref{A2} is approximated as a shift up by $2\sqrt{2}$ from the wavenumber in Eq.~\eqref{kms}, corresponding to  the Gaussian wavepacket approximation, to the wavenumber in Eq.~\eqref{alpha1}, corresponding to the local curvature of the Gaussian wavepacket approximation. Therefore, the red line in Fig.~\ref{k4} corresponds to Eq.~\eqref{alpha1}. This approximation agrees reasonably well the the fft transform of the local curvature of the numerically evaluated eigenfunctions, with a small systematic difference.

In Appendix~\ref{sec:bead_model}, we compare the local curvature of the elastica with the local curvature of an elastic fiber made of beads, taking into account the non-zero thickness and 
hydrodynamic interactions between all the beads by the multipole method \cite{Cichocki1999,Ekiel2009}.

\section{Conclusions}\label{sec:conc}

We analyzed the stability of small deformations of a slender elastic fiber in the shear flow \eqref{shear} of a very viscous fluid by solving the spectral problem for the linearized elastica equations. The infinitely thin fiber is close to straight at an arbitrary 3D orientation, with the perturbations in two perpendicular directions. Even though the spectral differential equations for both perturbations are coupled with each other, 
we showed that patterns of 3D buckling of elastica are described by one ordinary differential equation with a single parameter $\tilde{\chi}$. 
It happens that this equation is the same as found by \cite{chakrabarti2020} for the pure compressional flow but with shifted and rescaled eigenvalues and eigenfunctions.  We analyzed the eigenvalues and eigenfunctions for the odd and even modes. 

For very flexible fibers, with $\tilde{\chi}\gtrsim 1.5\cdot 10^4$, we proposed a wave-packet approximation to the most unstable eigenfunctions and shown that it is very accurate. We derived the eigenvalues as linear functions of $\tilde{\chi}$ and the wave-packet parameters as linear functions of $\sqrt{\tilde{\chi}}$. 

We also analyzed the local curvature of the eigenfunctions, determined numerically by the Chebyshev spectral method \cite{liu2022,repository_2023}. By taking the  fast Fourier transform of it, we demonstrated the linear scaling of the characteristic wavenumber with  $\sqrt{\tilde{\chi}}$. 

The scaling of the buckled shapes and of the instability-growth time derived here might be used to predict and describe the buckling of elastic microfilaments in fluid flows \cite{young2007stretch,wandersman2010,guglielmini2012,kantsler2012,manikantan2015,quennouz2015,hall,kanchan2019,marchioli2023slender}, with possible new applications.

{\small \;\\
\noindent {\bf Acknowledgements}. 
We thank Professor Howard A. Stone for helpful discussions and Professor Saintillan for a useful remark.\\
\par

\noindent {\bf Funding}. 
P.Z. and M.E.J.
were supported in part by the National Science Centre under
grant UMO-2021/41/B/ST8/04474. \\
\par

\noindent {\bf Author ORCIDs}.\\
Pawe\l\ Sznajder~0000-0002-6916-0884,\\
Lujia Liu~0000-0001-6002-2672,\\ Piotr Zdybel~0000-0001-7484-1425,\\ Maria Ekiel-Je\.zewska~0000-0003-3134-460X }

\appendix
\section{Motivation of the scaling based on the asymptotic expansion}\label{A1}
We can benefit from an approximated fiber shape as a wave packet to gain some understanding of the scaling presented in the previous subsection. Without loss of generality, we restrict ourselves to the odd approximated solutions from Eq. \eqref{post}. We are interested in a situation where $s$ is small and $\tilde{\chi}$ is simultaneously large.

We substitute $\Phi = \sin(ks)\exp(\xi s^2)$ to the left side of Eq. \eqref{unieq} and expand the result in powers of $s$. The left-hand side of Eq. \eqref{unieq} obtained during this procedure has the form: \bee \hspace{-0.cm}&& s (-k^5+20k^3\xi-60k\xi^2+k\tilde{\chi}+\frac{k^3\tilde{\chi}}{16}-\frac{3k\xi\tilde{\chi}}{8}-k\tilde{\sigma}\tilde{\chi}) \nonumber \\\hspace{-0.cm}&&+s^3(\frac{k^7}{6}-7k^5\xi+70k^3\xi^2
-140k\xi^3-\frac{3 k^3 \tilde{\chi}}{4}-\frac{k^5\tilde{\chi}}{96}+\frac{9k\xi\tilde{\chi}}{2}\nonumber
\\\hspace{-0.cm}&& +\frac{5k^3\xi\tilde{\chi}}{8}+\frac{k^3\tilde{\sigma}\tilde{\chi}}{6}-k\xi\tilde{\sigma}\tilde{\chi})+\mathcal{O}(s^5).\label{LHS} \eee  Moreover, we postulate that $k$, $\xi$ and $\tilde{\sigma}$ have the following dependence on $\tilde{\chi}\gg 1$ (this assumption is guided by the numerical fits in 
Eqs \eqref{fitk}-
\eqref{fitsigma}), 
\beq
k\!=\!\!\sum_{j=1}^{\mathcal{N}}K_j \tilde{\chi}^{(2-j)/2}\!,\;\;\;\;
\xi\!=\!\!\sum_{j=1}^{\mathcal{N}}M_j \tilde{\chi}^{(2-j)/2}\!, \;\;\;\;
\tilde{\sigma}\!=\!\!\sum_{j=1}^{\mathcal{N}}S_j \tilde{\chi}^{(3-j)/2}\!,
\label{exp}
\eeq
where $\mathcal{N}$ is the integer at which we truncate the expansions. In the resulting expansion of the expression \eqref{LHS}, the leading terms in $\tilde{\chi}$ for the given power of $s$ are proportional to $\tilde{\chi}^{(2n+3)/2} s^{2n-1}$, where $n=1,2,  \dots$. On the other hand, subleading terms are analogously commensurate with $\tilde{\chi}^{(2n+3-m)/2} s^{2n-1}$, where $m= 1, 2, \dots, m_{\mathrm{max}}$  is indexing consecutive terms in the expansion in powers of $\tilde{\chi}^{-1/2}$. The parameter $m_{\mathrm{max}}$ grows with values of $n$ and $N$, see 
\eqref{LHS} and \eqref{exp}. 

We illustrate the procedure of finding values of coefficients in Eq. \eqref{exp} for the truncation at $\mathcal{N}\!=\!1$. The terms of the expansion for $\mathcal{N}\!=\!1$ 
are schematically presented  in Table~\ref{ta1}, e.g. a cell for $n=2$ and $m=1$ represents the term  
for order $s^3\tilde{\chi}^3$ and similarly we interpret other terms. 
\begin{table}[h!]
\caption{Terms in the expansion \eqref{exp} of the expression \eqref{LHS}.}\label{ta1}
\begin{tabular}{|c|c|ccccc|}
\cline{3-7}
\multicolumn{2}{c|}{} & $m=0$ & $m=1$ & $m=2$  &   $m=3$ & \dots \\
\hline
$n$=1&$s^1$ & \cellcolor{green!10} $\tilde{\chi}^{5/2}$ & \cellcolor{green!10} $\tilde{\chi}^{2}$  &  \cellcolor{green!10} $\tilde{\chi}^{3/2}$  & \phantom{1} & \phantom{1} \\
$n$=2&$s^3$ & \cellcolor{blue!10} $\tilde{\chi}^{7/2}$ & \cellcolor{red!10} $\tilde{\chi}^{3}$ & \cellcolor{red!10}$\tilde{\chi}^{5/2}$ & \cellcolor{red!10}$\tilde{\chi}^{2}$ &  \phantom{1} \\
$n$=3& $s^5$ & \cellcolor{blue!10} $\tilde{\chi}^{9/2}$ & \cellcolor{red!10}$\tilde{\chi}^{4}$ & \cellcolor{red!10} $\tilde{\chi}^{7/2}$ & \cellcolor{red!10} $\tilde{\chi}^{3}$ & \cellcolor{red!10} $\tilde{\chi}^{5/2}$  \\
$\vdots$ &$\vdots$ & \cellcolor{blue!10}$\vdots$ & \cellcolor{red!10} $\vdots$  & 
\cellcolor{red!10} $\vdots$ &  \cellcolor{red!10}$\vdots$ & \cellcolor{red!10} $\ddots$\\
\hline
\end{tabular}
\end{table} 

We must find the values of  $K_1$, $M_1$, and $S_1$. Thus, we demand that coefficient for terms of leading orders $s^1\tilde{\chi}^{5/2}$, $s^1\tilde{\chi}^{2}$ and $s^1\tilde{\chi}^{3/2}$ (green) should identically be equal to zero, which leads to values of  $K_1=3/160$, $M_1=\sqrt{15}/30$ and $S_1=14319/655360000$. When the equation for order $s^1\tilde{\chi}^{5/2}$ is satisfied, then the dominant contributions for higher powers of $s$, i.e., $s^3 \tilde{\chi}^{7/2}$, 
$s^5\tilde{\chi}^{9/2}$, $\dots$ (violet) are vanishing. This is true for any value of $\mathcal{N}$ because these coefficients have the same form as for order $s^1\tilde{\chi}^{5/2}$ multiplied by a number and some power of $K_1$. Moreover, when we restrict to $s=o(\tilde{\chi}^{-1})$ (see, e.g., \cite{jeffreys1999methods}), all other terms in the expansion (pink) asymptotically go to zero. 

Similarly, one can obtain more accurate results for truncation at larger values of $\mathcal{N}$, but it complicates calculations. A higher value of $\mathcal{N}$ allows the application of less restrictive requirements for the asymptotic behavior of $s$, i.e. $s=o(\tilde{\chi}^{-\zeta})$, where $\zeta<1$.  We take $\mathcal{N}=3$ and obtain the following expressions for $k$, $\xi$, and $\tilde{\sigma}$:
\begin{align}
&k=\frac{\sqrt{2}}{8}\tilde{\chi}^{1/2}+\frac{\sqrt{2}}{2}+\mathcal{O}(\tilde{\chi}^{-1/2}), \nonumber \\
&\xi=-\frac{1}{8}\tilde{\chi}^{1/2}+0.0756+\mathcal{O}(\tilde{\chi}^{-1/2}), \nonumber
\\&
\tilde{\sigma}=\frac{1}{2^{10}}\tilde{\chi}-\frac{1}{2^{5}}\tilde{\chi}^{1/2} -0.6061+\mathcal{O}(\tilde{\chi}^{-1/2}).
\label{kms}
\end{align}
 For $\mathcal{N}$=3, we assume that $s=o(\tilde{\chi}^{-3/5})$, which guarantees that higher order terms in expansion in powers of $s$ and $\tilde{\chi}$ asymptotically vanish. As one can see by comparing Eqs \eqref{fitk}-\eqref{fitsigma} with Eq. \eqref{kms}, our analytical coefficients in Eq. \eqref{kms} are approximated with an excellent accuracy by the coefficients $K_1$, $M_1$, $S_1$ and $S_2$ obtained from the numerical fit in Eqs \eqref{fitk}-\eqref{fitsigma}. 

The asymptotic expansion presented above works for small values of $s$. However, the numerical fit demonstrates that the wavepacket  \eqref{post} with the parameters \eqref{kms} well approximates the eigenfunctions and eigenvalues in the whole range of $s$.

\section{Wavenumbers based on shape and curvature}\label{A2} 
The Fourier transform $F(\alpha)$ of  a function $f(s)$ is defined as
\beq
F_f(\alpha)=\int_{-\infty}^{+\infty} f(s) \exp^{-is\alpha} ds.
\eeq

For the wavepacket $\Phi(s)$ given by Eqs.~\eqref{post} and \eqref{kms}, the Fourier transform scales as
\beq
F_{\Phi}(\alpha) \propto \exp[(\alpha-k)^2/(4\xi)],\eeq
with the maximum of $F_{\Phi}(\alpha)$ at 
\beq\alpha=\alpha_0=k\approx \frac{\sqrt{2}}{8}(\sqrt{\tilde{\chi}}+4).
\eeq

The Fourier transform of the curvature, approximated as $\kappa'(s)\propto -d^2\Phi(s)/ds^2$, scales as
\beq
F_{\kappa'}(\alpha) \propto \alpha^2 \exp[(\alpha-k)^2/(4\xi)].\eeq
Calculating zeros of the derivative, and using Eqs.~\eqref{kms}, we determine that
the maximum of $F_{\kappa'}(\alpha)$
is located at \bee
&&\alpha =\alpha_1\approx  
\frac{k}{2}\left(1+\sqrt{1-\frac{16\xi}{k^2}}\;\right) \approx k-\frac{4\xi}{k} \approx \alpha_0 +2\sqrt{2} \nonumber \\&&\approx \frac{\sqrt{2}}{8}\sqrt{\tilde{\chi}} +
\frac{5\sqrt{2}}{2}.\label{alpha1}
\eee

In Eq.~\eqref{alpha1}, we demonstrated that the wavenumbers $\alpha_0$ and $\alpha_1$, based on shape and curvature, respectively, are not the same, and $\alpha_1>\alpha_0$. Within the adopted approximation, they are both linear functions of $\sqrt{\tilde{\chi}}$ with the same slope, but shifted with respect to each other. A more precise relation would be obtained if terms $\propto 1/\sqrt{\tilde{\chi}}$ were taken into account. In this case, $\alpha_1$ would be  a nonlinear function of $\sqrt{\tilde{\chi}}$. 

The curvature $\kappa$ used earlier is related to $\kappa'$ by the equation $\kappa\!=\!|\kappa'|$. Therefore, the characteristic wavenumber for $\kappa$ is $2\alpha_1$. 

\section{Evolution of an elastic filament with a non-negligible thickness}\label{sec:bead_model}
In this Appendix, we compare the 
local curvature of the elastica eigenfunctions with the local curvature of an elastic fiber  
with a non-negligible thickness, taking into account hydrodynamic interactions between all the fiber segments in the shear flow.

Following previous publications \cite{delmotte2015,gauger2006,Slowicka2015,slowicka2022} we use the bead model. An elastic filament of thickness $d$ is modeled as a chain of $N=40$ identical, spherical, solid beads with diameter $d$. Consecutive beads are connected by springs
with the spring constant $\mathcal{K}$, 
and the equilibrium distance between the bead centers (the bond length) is equal to $\ell_0=1.02d$.  Consecutive bonds are also connected by springs, with the bending stiffness  
$
\mathcal{A}\!=\! E \pi d^4/64
$, where $E$ is Young's modulus.
At the elastic equilibrium, the fiber is straight.
The stretching and 
bending potential energies of the filament are 
 \beq
E_s\!=\!\frac{\mathcal{K}}{2}
\sum_{i=2}^N (\ell_i\!-\!\ell_0)^2, \hspace{1cm} E_b=\!=\!\frac{\mathcal{A}}{2\ell_0}\sum_{i=2}^{N-1} \beta_i^2,\hspace{0.4cm}
\eeq
where ${\ell}_i\!=\!|\bm{\mathcal{R}}_i\!-\!\bm{\mathcal{R}}_{i-1}|$, 
$\cos \beta_i\!=\!(\bm{\mathcal{R}}_i\!-\!\bm{\mathcal{R}}_{i-1})\cdot(\bm{\mathcal{R}}_{i+1}\!-\!\bm{\mathcal{R}}_{i})/(\ell_i \ell_{i+1})$,
and    $\bm{\mathcal{R}}_i$ is the position of the 
center of bead $i$. 
With $d$ and $1/\dot{\gamma}$ as the length and time units, respectively, the dimensionless quantities are: 
$\bm{r}_i\!=\!\bm{\mathcal{R}}_i/d$, 
${k}_0\!=\!{\mathcal{K}}/{(\pi \mu d\dot{\gamma})}\mbox{ and }A\!=\!E/(64 \mu \dot{\gamma})$. The filaments are almost inextensible, $k_0\!=\!1000$, and their bending stiffness ratio $A\in [20, 1000]$.

Dynamics of the no-slip beads are evaluated from the Stokes equations by the multipole expansion, as in \cite{Felderhof1988,Kim1991}, corrected for lubrication and implemented in the precise {\sc Hydromultipole} numerical codes, described in \cite{Cichocki1999,Ekiel2009}.
As in \cite{Slowicka2015,Slowicka2020,slowicka2022,zuk2021universal},
positions $\bm{r}_i$ of the bead centers satisfy the following equations,  \begin{equation} \label{eq:velocity}
 \dot{\bm{r}}_{i} - \bm{v}_0(\bm{r}_i) = \sum_{j=1}^N \left(
                     \bm{\mu}^{tt}_{ij} \cdot \bm{F}_j
                   + \bm{\mu}^{td}_{ij} : \bm{E}_{\infty}
                 \right),\hspace{0.4cm}i=1,...,N,
\end{equation}
where $\bm{F}_j=\frac{\partial}{\partial \bm{r}_j} (E_s+E_b)/(\pi \mu d^3\dot{\gamma})$
is the dimensionless elastic force acting on the bead $j$, and
$\mathbf{E}_{\infty}=\frac{1}{2}(\bm{\nabla} \bm{v}_0+ \left(\bm{\nabla} \bm{v}_0 \right)^{T} ) $ is the rate-of-strain tensor.
The mobility matrices, $\bm{\mu}^{tt}_{ij}$ and $\bm{\mu}^{td}_{ij}$, depend on the positions of all the bead centers and are evaluated by the {\sc Hydromultipole} program, with the multipole truncation order $L=2$. In Ref.~\cite{Slowicka2015} the approximation of $L=2$ was compared with the approximations of $L=3$ and $L=4$, and it was shown that the difference is of the order of 2\%, and therefore $L=2$ is sufficiently accurate (while much faster than higher values of $L$).

In Ref.~\cite{slowicka2022} numerical simulations were performed of initially straight elastic fibers, and obtained only odd buckled shapes. It is caused by the fiber thickness which promotes an odd deformation. To observe even shapes, an even initial perturbation is needed, as in the numerical simulation in Ref.~\cite{liu2018}.  Therefore, 
in this paper, we initially impose a small, but significant either odd or even perturbation   of the center of bead $i$, relative to the center of a straight fiber in elastic equilibrium, oriented at $\theta_0\!=\!90^{\circ} \mbox{ and } 
\phi_0\!=\!160^{\circ}$. 
The  perturbation does not depend on $A$ and is restricted to the shear plane, $\delta_i (-\sin \phi_0,  \cos \phi_0, 0)$, with  
$\delta_i\!=\! a\sum_{m=1}^{5}(-1)^{q_m}\sin[2m\pi(\frac{i-1}{N-1}-\frac{1}{2})/(2m)^2$ for the odd perturbation,
where $q_m$ are independent random variables equal to 1 or 2 with probability $1/2$. Finally, the positions are shifted to make all the distances between the bead centers equal to the equilibrium value $\ell_0/d=1.02$.
The value of $a$ is chosen so that the maximum distance of a bead center from the unperturbed filament is $d_m \approx 0.076$. 
The even initial perturbation is constructed similarly, with 
$d_m \approx 0.19$. Both initial perturbations of all the beads are listed in an open repository \cite{initSZLE}. 

We compare an elastic filament of length $L$ and diameter $d$, made of $N=40$ beads, to the elastica with $L/d\approx 40$, 
and the same  
value of Young's modulus $E$.  
Therefore, Eq.~\eqref{def eta} and the relation $A=E/(64 \mu \dot{\gamma})$ determine $\eta$ as a function of $A$,
\bee
\eta=
\frac{2}{\ln(2L/d)}\left(\frac{L}{d}
\right)^4 A^{-1}=1.17\cdot 10^6 A^{-1}.\label{etaA}
\eee

In the simulations, for each $A$ we select the filament shape at time $t=0.75$, sufficient for the development of the buckling instability and small enough for a small change of the angle $\phi$ of the filament principal axis, 
as shown in Ref.~\cite{slowicka2022} in Fig.~13. Then, we compare it 
with the most unstable eigenfunction of the same parity, $\Phi_{u1}(\tilde{\chi})$, where 
\beq
\tilde{\chi}=
-\eta \sin(2\phi_0) \sin^2\theta_0=7.5\cdot10^5 A^{-1}.
\eeq

To analyze the filament shapes, 
we use the local curvature. 
At the center of bead $i\!=\!2,...,N\!-\!1$ it
is given as
\beq
\kappa_{i}=\frac{2\left|\left(\bm{r}_{i-1}-\bm{r}_{i}\right)\times\left(\bm{r}_{i}-\bm{r}_{i+1}\right)\right|}{|\bm{r}_{i-1}-\bm{r}_{i}||\bm{r}_{i}-\bm{r}_{i+1}||\bm{r}_{i+1}-\bm{r}_{i-1}|}.
\label{kappa_i}
\eeq
We use  
$s_i=-0.5+(i-1)/(N-1)$ as 
a discrete analog of the arclength $s$. Then, $s_i$ and $\kappa_i$ are interpolated and $\kappa_i$  is normalized by its maximum value along the filament.

\begin{figure}[h!tbp]
  \includegraphics[width=0.45\textwidth]{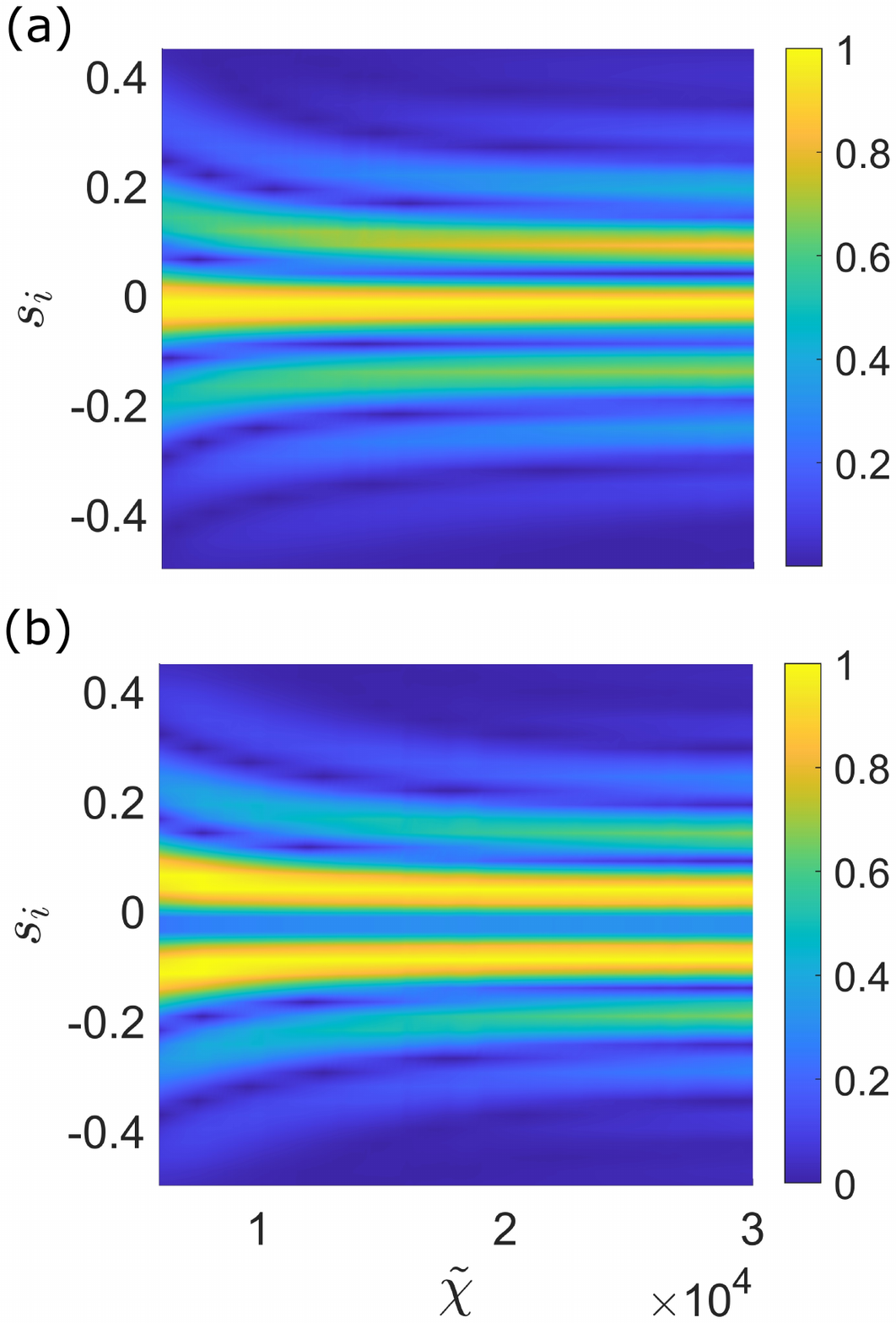}\vspace{-.5cm}
  \caption{Local curvature of an elastic filament, normalized by its maximum value along the filament and shown as colors.  Numerical simulations for the elastic filament made of beads, at $t\!=\!0.75$ for initially almost straight configuration oriented at the angles $\theta_0\!=\!90^{\circ}$ and $\phi_0\!=\!160^{\circ}$, with even (a), and odd (b), initial perturbations.}\label{s4}
\end{figure}

The local curvature as a function of $s_i$ and $\tilde{\chi}$ is shown in colors in Fig. \ref{s4}, 
separately for the even and odd initial perturbations. We observe that at $t=0.75$, the 
parity of the filament shapes is the same as the parity of the initial perturbation. The dependence of the fiber shape on $\tilde{\chi}$, shown in Fig. \ref{s4}, is very similar as in the case of the elastica, shown in Fig.~\ref{c4}. 
For large values of $\tilde{\chi}$, both models agree with each other qualitatively. 
We observe the increasing number of the local maxima of the fiber curvature for the increasing value of $\tilde{\chi}$, as in Fig.~12 in Ref.~\cite{slowicka2022}, corresponding to the increase of the characteristic wavenumber. For large values of $\tilde{\chi}$, the increase of the wavenumber of an elastic filament made of beads is slightly slower than  of the elastica.

\end{document}